\documentclass[twocolumn,showpacs,showkeys,aps,preprintnumbers]{revtex4}

\usepackage{graphicx}
\usepackage{dcolumn}
\usepackage{bm,mathrsfs,amsmath,amssymb,braket}
\usepackage{pst-all}
\psset{unit=1cm,fillcolor=white,linewidth=.3mm,arrowsize=5pt,coilarm=.25cm,coilwidth=.25cm}
\DeclareRobustCommand\openone{\leavevmode\hbox{\small1\normalsize\kern-.33em1}}%
\newcommand{\tr}[2][\mbox{\!}]{\textrm{tr}_{#1}\left[#2\right]}%
\newcommand{\definido}{\stackrel{\scriptscriptstyle\textrm{def}}{\displaystyle =}}%
\newcommand{\vi}{\hat{\imath}}
\newcommand{\vj}{\hat{\jmath}}
\newcommand{\vk}{\hat{k}}
\renewcommand{\Re}{\textnormal{Re}}

\DeclareRobustCommand{\exponencial}{\leavevmode\hbox{\LARGE\textnormal{e}\normalsize}}
\renewcommand{\exp}{\exponencial}
\newrgbcolor{brown}{0.50 0.25 0.00}

\begin{document}
\preprint{APS/123-QED}
\title{Exact analytical solution of Quantum Discord for Generalized Werner-Like Non-X states}%
\author{S. D\'{i}az-Sol\'{o}rzano}%
\author{E. Castro}
\altaffiliation[Electronic address:\space]{\texttt{ecastro@usb.ve}}%
\affiliation{%
Physics Department, Quantum Information and Communication Group, Sim\'{o}n Bol\'{i}var
University, Apdo. 89000, Caracas 1086, Venezuela.
}%

\date{\today}

\begin{abstract}\noindent%
The Generalized Werner-Like states (GWLs) are a class of non--X states in which
the exchange operator is replaced for a generic one-rank projector in the
Werner states. We obtained an exact analytical expression of Quantum Discord
for these states. The optimization problem involved is solved by giving an
analytical expression, in exact form for the conditional entropy. We compared
the Quantum Discord (QD) with the Entanglement of Formation (EoF) for the same
states. The pure states of GWLs with equal concurrence have the same QD and
EoF.
\end{abstract}

\pacs{
      03.65.Ud,\space
      03.67.-a,\space
      03.65.-w.\space
      }%
\keywords{Quantum Discord, Entangled of Formation, Werner states, Werner-Like states.}%
\maketitle%
\section{Introduction}
\label{sec:intro}%
Quantum correlations lie in the foundation of quantum mechanics and are the heart of quantum
information science. They are important to study the differences between the classical and quantum
worlds because, in general, the quantum systems can be correlated in ways inaccessible to classical
objects. The research on quantum correlation measures were initially developed on the
entanglement-separability paradigm~\cite{Amico} (and the references therein). However, it is well
known that entanglement does not account for all quantum correlations and that even correlations of
separable states are not completely classical. Entanglement is an inevitable feature of not only
quantum theory but also any non-classical theory~\cite{RichensSelbyAl-Safi}, and this is necessary
for emergent classicality in all physical theories. The study of quantum correlation quantifiers
other than entanglement, such as the Quantum Discord (QD), has a crucial importance for the full
development of new quantum technologies because it is more robust than entanglement against the
effects of decoherence~\cite{Werlang,Modi,Yune,Wang}, and can be among others a resource in quantum
computation~\cite{Dattaa,Dattab,Ke}, quantum non-locality~\cite{Liu}, quantum key
distribution~\cite{Brodutch}, remote state preparation~\cite{Dakic}, quantum
cryptography~\cite{Pirandola} and quantum coherence~\cite{Yao}.
\\
\par\indent%
Experimentally it is difficult to prepare pure states, the researcher must do a thorough
examination of the system to know the possible pure states to which the system can access. In
general, the states are mixed since they characterize the interaction of the system with its
surrounding environment. The study of the quantum information properties of mixed states is more
complicated and less understood that the pure states. The set of Werner states~\cite{Wernera} (Ws)
is an important type of mixed states, derived in 1989, which plays a fundamental role in the
foundations of quantum mechanic and quantum information theory. Since these states admit a hidden
variable model without violating Bell's inequalities, then the correlation measured that are
generated with these states can also be described by a local model, despite of being entangled.
Moreover, these states are used as quantum channels with noise that do not maintain the additivity,
they are also employed in the study of deterministic purifications~\cite{Lyons2012}. On the other
hand, the \emph{Werner-Popescu states}~\cite{Popescu-PRL.72.797} (WPs), the \emph{Quasi-Werner
states}~\cite{Horodecki-PhysRevA.59.4206} (QWs) and the \emph{Bell Werner-Like
state}~\cite{Horodeckia} (BWLs), also called noise singlets, for bipartite system of qubits, are
mixing states maximally entangled and have been studied widely as a fundamental resource for the
quantum information processing, and also in the study of non-local properties in quantum mechanics.
These mixing states are the most natural generalization of the GWLs. The GWLs (for detail see
section~\ref{sec:GWLs}) are a family of mixed states, obtained by the convex sum between a
maximally mixed state (also called unpolarized state) and a one-rank projector built with a
Generalized Bipartite pure state (GBps).
\\
\par\indent%
The QD, as a quantum correlation of a bipartite system, initially introduced by Zurek and
Olliver~\cite{Zurek,Ollivier} and by Henderson and Vedral~\cite{Henderson}, is a more general
concept to measure quantum correlations than quantum entanglement, since separable mixed states can
have nonzero QD. This measures the fraction of the pairwise mutual information that is locally
inaccessible in a multipartite system (for detail see section~\ref{sec:QDGWLs}). The QD is also
called the \emph{locally inaccessible information} (LII)~\cite{Fanchinib}, since the QD measured on
one partition is the information of the system that is inaccessible to an observer in other
partition. In this context, quantum measurements only provide information on the partition
measured, however, simultaneously they introduce disturbance and destroy the coherence in the
system. One of the problems QD has is the minimization process involved for the calculation of the
conditional entropy. Until now, the QD only has been obtained for a few special classes of
two-qubit X--states~\cite{LuoPRA77.042303.2008,Ali,Maldonado,Yurischev}, and generally this is
determined numerically~\cite{Girolami}. Yao and collaborators~\cite{Yao.Dong.Zhu} has evaluated
numerically the QD for a special class of non--X states when the Bloch vectors are orthogonal
vectors. This class of states cannot be written as a GWLs, since in the representation of
Fano-Bloch both states do not match. Recently, Huang \cite{Huang} obtained a precise mathematical
characterization of the computational difficulty of EoF and QD. In particular, he proved that
computing a large class of entanglement measures (including, but not limited to, EoF) and computing
QD are NP-complete and NP-hard in some particular cases. The QD is not always larger than the
entanglement, and there is not clear evidence of the relationship between entanglement and quantum
discord~\cite{Luob,Li}, in general, since they seem to capture different properties of the states.
The principal aim of this paper is to derive analytical solutions of QD for the GWLs built with
GBps, and compare the QD with a measure of entanglement, specifically the EoF.
\\
\par\indent%
This paper is organized as follows. A detailed review of the GWLs is given in Sec.~\ref{sec:GWLs}.
In Sec.~\ref{sec:QDGWLs} we present an analytical approach to obtain the exact solutions of the QD
for the GWLs, while in Sec.~\ref{sec:EoFGWLs} we determine the EoF for GWLs. In
Sec.~\ref{sec:examples} we evaluate the QD and EoF for several states GBps, using four discrete
state with different concurrences, here show the behavior of the EoF and QD with mixing parameter.
In this section, also we prove that the QD is a monotonous function of the concurrence of the GBps.
Finally, in the Secs.~\ref{sec:Result} and \ref{sec:Conclusion} we present the analysis drawn from
our results and conclusions of work. Additionally, three appendices are included which contain the
calculation of the projective measure on pure state, the calculation of the conditional entropy for
the GWLs and the calculation of the critical point for mixing parameter where Bell's inequality is
violated.
\section{Generalized Werner-Like non--X states}
\label{sec:GWLs}%
Let $\ket{ij}$ be the computational bases in space $\mathscr{H}_2\otimes\mathscr{H}_2$, where
$\set{ij}=\{00,01,10,11\}$ and $\mathscr{H}_2$ is the Hilbert space of dimension two. The GBps
$\ket{\psi}$ is given by,
\begin{equation}\label{eq:27}
\ket{\psi}=z_1\ket{00}+z_2\ket{01}+z_3\ket{10}+z_4\ket{11}.
\end{equation}
The complex numbers $z_i$ (with $i=1,2,3,4$) are those that verify the normalization condition
$\sum_i|z_i|^2=1$. The GBps can be represented by a $2\times2$ matrix whose elements are obtained
with the components of the pure satate \eqref{eq:27}, in accordance with
\vspace{-.4cm}%
\begin{equation}\label{eq:2}
\widehat{\mathbb{W}}_\psi\definido[\braket{ij|\psi}]_{2\times2}=%
\begin{bmatrix}
\braket{00|\psi} & \braket{01|\psi} \\*[.1cm]
\braket{10|\psi} & \braket{11|\psi} \\
\end{bmatrix}=
\begin{bmatrix}
  z_1 & z_2 \\*[.1cm]
  z_3 & z_4 \\
\end{bmatrix}%
\end{equation}
\\*[-.8cm]
\par\noindent%
The normalization condition of GBps $\ket{\psi}$ in term of matrix $\widehat{\mathbb{W}}_\psi$ it
is written as $\tr{\widehat{\mathbb{W}}_\psi\widehat{\mathbb{W}}_\psi^\dag}=1$ (see
appendix~\ref{sec:apenA} for details). The one-rank projector built with the GBps $\ket{\psi}$ or
the density matrix of the GBps $\ket{\psi}$ is denoted as
\begin{equation}\label{eq:32}
\hat{\mathbb{P}}_\psi\definido\ket{\psi}\bra{\psi}=%
\begin{bmatrix}
|z_1|^2      & z_1\bar{z}_2 & z_1\bar{z}_3  & z_1\bar{z}_4\\
z_2\bar{z}_1 & |z_1|^2      & z_2\bar{z}_3  & z_2\bar{z}_4\\
z_3\bar{z}_1 & z_3\bar{z}_2 & |z_3|^2       & z_3\bar{z}_4\\
z_4\bar{z}_1 & z_4\bar{z}_2 & z_4\bar{z}_3  & |z_4|^2      \\
\end{bmatrix}.
\end{equation}
The $\bar{z}_i$ shown in the expression \eqref{eq:32} are the conjugate complex of $z_i$, with
$i=1,2,3,4$.
\\
\par\indent%
On the other hand, the bipartite Ws of qubits are self-adjoint operators, bounded and of class
trace that act onto the composite space $\mathscr{H}_2\otimes\mathscr{H}_2$, formed by
\begin{equation}\label{eq:0}
\begin{split}
\rho_{\textrm{W}}(p)%
&=\tfrac{1-p}{4}\hat{\openone}_{4}+\tfrac{p}{2}\hat{\mathbb{F}}_4=%
  \tfrac{1+p}{4}\hat{\openone}_{4}-p\hat{\mathbb{P}}_{\Phi_-},
\\%
&=\frac{1}{4}\begin{bmatrix}
1+p&   0&  0 &  0\\
0  & 1-p&  2p&  0\\
0  &  2p& 1-p&  0\\
0  &   0&  0 &  1+p\\
\end{bmatrix},
\end{split}
\end{equation}
being $\ket{\Phi_-}=\tfrac{1}{\sqrt{2}}(\ket{01}-\ket{10})$ a Bell state, and $p$ is the mixing
parameter with $p\in\left[-1,\tfrac{1}{3}\right]$. Its range of values guarantees the positivity of
the Ws. Furthermore, $\hat{\mathbb{F}}_4$ is the exchange operator defined by
\begin{equation}\label{eq:51}
\hat{\mathbb{F}}_4=
\sum\limits_{i,j=0}^{3}\ket{ij}\bra{ji}=%
\hat{\openone}_{4}-2\hat{\mathbb{P}}_{\Phi_-}=%
\left[\begin{smallmatrix}
  1 & 0 & 0 & 0 \\
  0 & 0 & 1 & 0 \\
  0 & 1 & 0 & 0 \\
  0 & 0 & 0 & 1 \\
\end{smallmatrix}\right],
\end{equation}
The Ws, given in the expression \eqref{eq:0}, are states--X invariant under any local unitary
operator of the form $\hat{\mathbb{U}}\otimes\hat{\mathbb{U}}$, they admit a model of hidden
variables~\cite{Wernera} if $p\in[-1,-\tfrac{1}{2}]$, being still entangled.
\\
\par\indent%
The GWLs are a one-parametric family of mixed states, obtained by the convex sum between a
maximally mixed state (also called unpolarized state) and a one-rank projector built with the GBps,
given by expression \eqref{eq:27}. In other words, the GWLs is an generalization of the Ws when
exchange the operator $\tfrac{1}{2}\hat{\mathbb{F}}$ by the projector $\hat{\mathbb{P}}_\psi$, and
furthermore exchance the mixing parameter $p$ by $-p$. Then, the density matrix of fourth orden for
the GWLs has the form:
\begin{widetext}
\mbox{}\vspace{-.5cm}\mbox{}%
\begin{equation}\label{eq:4}
\rho_{\textrm{GWL}}(\psi,p)=%
\tfrac{1- p}{4}\hat{\openone}_{4}+p\hat{\mathbb{P}}_{\psi}=%
\begin{bmatrix}
\tfrac{1-p}{4}+p|z_1|^2&     pz_1\bar{z}_2     &      pz_1\bar{z}_3    &  pz_1\bar{z}_4 \\
     pz_2\bar{z}_1     &\tfrac{1-p}{4}+p|z_2|^2&      pz_2\bar{z}_3    &  pz_2\bar{z}_4 \\
     pz_3\bar{z}_1     &     pz_3\bar{z}_2     &\tfrac{1-p}{4}+p|z_3|^2&  pz_3\bar{z}_4 \\
     pz_4\bar{z}_1     &     pz_4\bar{z}_2     &      pz_4\bar{z}_3    &\tfrac{1-p}{4}+p|z_4|^2 \\
\end{bmatrix}.
\end{equation}
\end{widetext}
\mbox{}\\*[-1cm]
The range of variation of the mixing parameter $p$ is now $-\tfrac{1}{3}\leq p\leq1$, which
guarantees the positivity of the GWLs. The parameter $p$, considered in the expression
\eqref{eq:4}, is understood as a probability when the range of variation is $0\le p\le1$. In this
case the GWLs represents a convex sum of the density matrix of the GBps $\ket{\psi}$ and
non-coherent density matrix of an unpolarized state (white noise), with probabilities $p$ and
$1-p$, respectively. The WPs, QWs and BWLs are obtained where the Bell states
$\ket{\Psi_{\pm}}=\tfrac{1}{\sqrt{2}}(\ket{00}\pm\ket{11})$ and
$\ket{\Phi_{\pm}}=\tfrac{1}{\sqrt{2}}(\ket{01}\pm\ket{10})$ are used to built the projector
$\hat{\mathbb{P}}_\psi$ in the expression \eqref{eq:4}. One difference between the states
\eqref{eq:0} and \eqref{eq:4} is that the Ws are $X$ states while that the GWLs are not--X states
in general, unless $z_1=z_4=0$ or $z_2=z_3=0$; like the WPs, QWs and BWLs. Othes fundamental
difference is that $\hat{\mathbb{F}}$ is an involutive operator
($\hat{\mathbb{F}}^2=\hat{\openone}$) while $\hat{\mathbb{P}}_\psi$ is an idempotent operator
($\hat{\mathbb{P}}^2_\psi=\hat{\mathbb{P}}_\psi$), so that the GWLs generate in principle different
correlations that the Ws, since the replacing of $\tfrac{1}{2}\hat{\mathbb{F}}$ by
$\hat{\mathbb{P}}_{\psi}$ in the expression \eqref{eq:0} makes that $\rho_{\textrm{W}}(p)$ do not
be unitarily equivalent to $\rho_{\textrm{GWL}}(\psi,p)$, for this reason the GWLs loosses the
invariance under local unitary transformations. Nevertheless, the GWLs and Ws are connected by the
transformation
\begin{equation}\label{eq:53}
\rho_{\textrm{GWL}}(\Phi_{-},-p)=\tfrac{1+p}{4}\,\hat{\openone}_4-p\,\hat{\mathbb{P}}_{\Phi_{-}}
\equiv\rho_{\textrm{W}}(p).
\end{equation}
This equality is exact only in four dimensions. In other dimensions it is impossible to obtain the
equality \eqref{eq:53}. But any unitary transformation applied on GWLs leaves them invariant in
shape, without changing the mixing parameter, this is
\begin{equation}\label{eq:5}
\begin{split}
\rho_{\textrm{GWL}}(\psi,p)\xrightarrow{\hat{\mathbb{U}}}%
\hat{\mathbb{U}}\rho_{\textrm{GWL}}(\psi,p)\hat{\mathbb{U}}^\dag%
&=\tfrac{1-p}{4}\hat{\openone}_{4}+p\hat{\mathbb{P}}_{\psi_U}
\\%
&=\rho_{\textrm{GWL}}(\psi_U,p),
\end{split}
\end{equation}
where $\ket{\psi_U}=\hat{\mathbb{U}}\ket{\psi}$. The Ws changed by no local unitary transformations
are called \emph{Werner Derivative states} (WDs), and these states lead to a type of GWLs. The
study of local and nonlocal properties is done in reference~\cite{Hiroshima2000}, but this study is
incomplete since it only considers a particular class of unitary transformations. Therefore, all
the correlations contained in the WDs are present in the GWLs.
\section{Entanglement of formation for the GWLs}
\label{sec:EoFGWLs}%
A good measure to quantify the entanglement of a pure state $\ket{\psi}$ is the von-Neumann
entropy~\cite{vonNeumann-libro}, since a pure state can be constructed from a set of maximally
entangled singlet states and the number of these states is proportional to the entropy of the
reduced states of any partition~\cite{Woottersa,Woottersb}. However, the von-Neumann entropy is not
a good measure of the degree of entanglement for mixed states because there are product states
whose partitions may have entropies different from zero, for example, $\rho=\rho_1\otimes\rho_2$
with $S[\rho_1]\neq0$. In orden to quantify the degree of entanglement in arbitrary bipartite state
Wootters~\cite{Woottersa,Woottersb} proposed the EoF, given by
\begin{equation}\label{eq:33}
\textrm{EoF}[\rho]=H_2\left(\tfrac{1+\Delta_\rho}{2}\right),%
\;\;\textrm{with}\;\;\Delta_\rho\definido\sqrt{1-C^2[\rho]}.
\end{equation}
The function $H_2(z)\definido-z\log_2(z)-(1-z)\log_2(1-z)$ shown in the equation \eqref{eq:33} is
the Shannon binary entropy function, and $C[\rho]$ is the \emph{concurrence} function of the state
$\rho$, defined as
$C[\rho]\definido\max\{0,\sqrt{\lambda_1}-\sqrt{\lambda_2}-\sqrt{\lambda_3}-\sqrt{\lambda_4}\}$.
The $\lambda_i$'s are the eigenvalues of the positive operator $\rho\widetilde{\rho}$, arranged in
decreasing order. The operator $\widetilde{\rho}$ is the spin-flip operation on the conjugate of
the state $\rho$, i.e.
$\widetilde{\rho}=(\sigma_y\otimes\sigma_y)\overline{\rho}(\sigma_y\otimes\sigma_y)$, being
$\overline{\rho}$ the conjugate complex of $\rho$. In the case of a pure state $\ket{\psi}$, the
spin-flip operation onto the conjugate complex of the state is given by
$\widetilde{\rho}=(\sigma_y\otimes\sigma_y)\ket{\overline{\psi}}\bra{\overline{\psi}}(\sigma_y\otimes\sigma_y)\equiv\ket{\widetilde{\psi}}\bra{\widetilde{\psi}}$
so that $\rho\widetilde{\rho}=\braket{\psi|\widetilde{\psi}}\ket{\psi}\bra{\widetilde{\psi}}$, and
the characteristic equation $\rho\widetilde{\rho}\ket{\lambda}=\lambda\ket{\lambda}$ leads to
$\lambda=|\braket{\psi|\widetilde{\psi}}|^2$, after projecting this equation on
$\bra{\widetilde{\psi}}$. Also, the determinant of $\ket{\psi}\bra{\widetilde{\psi}}$ is zero and
therefore $\rho\widetilde{\rho}$ has a null eigenvalue with multiplicity three which corresponds to
the ortogonal projection to the state $\ket{\widetilde{\psi}}$. In this sense,
$\sqrt{\lambda_1}=|\braket{\psi|\widetilde{\psi}}|$ and $\lambda_2=\lambda_3=\lambda_4=0$, being
the concurrence for a pure state $\ket{\psi}$
\begin{equation}\label{eq:16}
\begin{split}
C[\ket{\psi}]%
&=|\braket{\psi|\widetilde{\psi}}|=|\bra{\psi}\sigma_{y}\otimes\sigma_{y}\ket{\overline{\psi}}|,
\\
&=2|z_1z_4-z_2z_3|\equiv2|\det\widehat{\mathbb{W}}_\psi|.
\end{split}
\end{equation}
With the aim of obtaining the EoF for the GWLs, and compare it with the QD of same state, we
calculate the Wootters concurrence given in the equation \eqref{eq:33}. For the case of GWLs, the
spin-flip operation applied on the conjugate complex of the states defined by equation \eqref{eq:4}
is
\begin{equation}\label{eq:35}%
\widetilde{\rho}_{\textrm{GWL}}(\psi,p)=%
\tfrac{1-p}{4}\hat{\openone}_4+p\ket{\widetilde{\psi}}\bra{\widetilde{\psi}}\equiv%
\rho_{\textrm{GWL}}(\tilde{\psi},p),
\end{equation}
while
\begin{equation}\label{eq:36}%
\rho_{\textrm{GWL}}(\psi,p)\widetilde{\rho}_{\textrm{GWL}}(\psi,p)=%
\left(\tfrac{1-p}{4}\right)^2\hat{\openone}_4+\hat{\mathbb{A}},%
\end{equation} where
\begin{small}\begin{equation}\label{eq:37}%
\hat{\mathbb{A}}=p^2C[\ket{\psi}]\exp^{i\phi}\ket{\psi}\bra{\widetilde{\psi}}+%
\tfrac{p(1-p)}{4}\left(\ket{\psi}\bra{\psi}+\ket{\widetilde{\psi}}\bra{\widetilde{\psi}}\right).
\end{equation}
\end{small}
\null\hspace{-.35cm} Here we have replaced $\braket{\psi|\widetilde{\psi}}$ by
$C[\ket{\psi}]\exp^{i\phi}$, where $\phi$ is the argument of $\braket{\psi|\widetilde{\psi}}$. On
the other hand, the eigenvectors of the matrix $\hat{\mathbb{A}}$ are equal to the eigenvectors of
$\rho_{\textrm{GWL}}(\psi,p)\widetilde{\rho}_{\textrm{GWL}}(\psi,p)$, so we will focus on finding
the eigenvalues of this matrix. It is clear from equation \eqref{eq:37} that the domain of
$\hat{\mathbb{A}}$ can be expanded as linear combinations of the pure states $\ket{\psi}$ and
$\ket{\widetilde{\psi}}$, which means that the eigenvectors of $\hat{\mathbb{A}}$ can be written as
$\ket{\lambda}=\Lambda_1\ket{\psi}+\Lambda_2\ket{\widetilde{\psi}}$. Projecting the equation of
eigenvectors $\hat{\mathbb{A}}\ket{\lambda}=\lambda\ket{\lambda}$ on the states $\ket{\psi}$ and
$\ket{\widetilde{\psi}}$ we obtain an equation system for $\braket{\psi|\lambda}$ and
$\braket{\tilde{\psi}|\lambda}$, from which a straightforward calculation yields
\begin{equation}\label{eq:39}%
\begin{bmatrix}
\tfrac{p(1-p)}{4}-\lambda   & \frac{p(1+3p)}{4}C[\ket{\psi}]\exp^{i\phi} \\*[.2cm]
\tfrac{p(1-p)}{4}C[\ket{\psi}]\exp^{-i\phi} & \tfrac{p(1-p)}{4}+p^2C^2[\ket{\psi}]-\lambda%
\end{bmatrix}\,
\begin{bmatrix}
\braket{\psi|\lambda}
\\*[.2cm]%
\braket{\widetilde{\psi}|\lambda}
\end{bmatrix}=0.
\end{equation}
To determine a solution other than the trivial one, we impose that the determinant of the equation
system is zero and obtain the following eigenvalue equation
\begin{equation}\label{eq:40}%
\lambda^2-\frac{p^2(1-2\Delta_{\ket{\psi}}^2)+p}{2}\lambda+\frac{p^2(1-p)^2}{16}\Delta_{\ket{\psi}}^2=0,
\end{equation}
From this equation, two eigenvalues are determined. The other two eigenvalues of the operator
$\hat{\mathbb{A}}$ that correspond to the eigenvectors expanded into
$\{\ket{\psi},\ket{\widetilde{\psi}}\}$ are zero because $\det(\hat{\mathbb{A}})=0$. Finally, the
eigenvalues of equation \eqref{eq:36} are in decreasing order
\begin{subequations}\label{eq:41}
\begin{eqnarray}
\label{eq:41a}%
&&\hspace{-1.1cm}\lambda_1\hspace{-.1cm}=\hspace{-.1cm}%
\left(\tfrac{1-p}{4}\right)^2\hspace{-.2cm}+\hspace{-.1cm}%
\tfrac{p(1-p+2pC^2[\psi])+|p|C[\ket{\psi}]\sqrt{(1+p)^2-4p^2\Delta_{\ket{\psi}}^2}}{4},
\\%
\label{eq:41b}%
&&\hspace{-1.1cm}\lambda_2\hspace{-.1cm}=\hspace{-.1cm}%
\left(\tfrac{1-p}{4}\right)^2\hspace{-.2cm}+\hspace{-.1cm}%
\tfrac{p(1-p+2pC^2[\psi])-|p|C[\ket{\psi}]\sqrt{(1+p)^2-4p^2\Delta_{\ket{\psi}}^2}}{4},
\end{eqnarray}
and
\begin{eqnarray}
\label{eq:41c}%
&&\hspace{-1cm}\lambda_3=\lambda_4=\left(\tfrac{1-p}{4}\right)^2,
\end{eqnarray}
\end{subequations}
so that the concurrence for GWLs is given by
\begin{equation}\label{eq:42}%
C[\rho_{\textrm{GWL}}]=\max\left\{0,\sqrt{\lambda_1}-\sqrt{\lambda_2}-\tfrac{1-p}{2}\right\}.
\end{equation}
Matching $\sqrt{\lambda_1}$ and $\sqrt{\lambda_2}+\tfrac{1-p}{2}$ we get the value of the largest
mixing parameter $p_c$ from which the EoF is zero, thereby the critical mixing parameter is given
by
\begin{equation}\label{eq:3}
p_c=\frac{1}{1+2C[\ket{\psi}]}.
\end{equation}
This quantiy is a critical value that limits the border between entanglement and separability of
GWLs. In other words, the GWLs are separable when $-\tfrac{1}{3}\le p\le p_c$ and entangled when
$p_c<p\le1$. So that, the critical value $p_c$ decreases monotonously with the increase of the
concurrence of the GBps $\ket{\psi}$. In particular, for BWLs o WPs we have the usual
result~\cite{Horodecki-PhysRevA.59.4206}, namely, they are entangled if $1/3<p\le1$ and classically
correlated if $-1/3\le p\le1/3$, since $\ket{\psi}$ is maximally entangled, i.e. $C[\psi]=1$. When
the pure state $\ket{\psi}$ is a product state ($C[\psi]=0$) then all the GWLs are a convex sum of
product states. In effect, taking $\ket{\psi}=\ket{A}\otimes\ket{B}$ result that
\begin{equation}\label{eq:9}
\rho_{\textrm{GWLs}}(A\otimes B,p)=%
\sum_{i,j=0}^{1}\hspace{-.1cm}\tfrac{1-p}{4}\hat{\mathbb{P}}_i\otimes\hat{\mathbb{P}}_j+%
p\hat{\mathbb{P}}_A\otimes\hat{\mathbb{P}}_B
\end{equation}
where $\hat{\mathbb{P}}_i\definido\ket{i}\bra{i}$ (with $i={0,1,A,B}$) for value of mixing
parameters where the GWLs are separable. However, not all the projectors in the convex sum are
orthogonal, so there would be correlations not necessarily classic.
\section{Quantum discord of GWLs}
\label{sec:QDGWLs}%
The fundamental amount for the study of quantum information, in terms of its uncertainty, is the
von-Neumann entropy~\cite{vonNeumann-libro}. Namely, when uncertainty grows the state contains less
information. According to Wilde~\cite{Wilde-libro} this quantity measures the expected value of
quantum information content. This quantity is defined in bits as
$S[\rho]\definido-\tr{\rho\log_{2}\rho}=-\sum_{i}\lambda_{i}\log_{2}\lambda_{i}$, where the
$\lambda_{i}$'s are the eigenvalues of the density operator $\rho$. For pure states
($\rho=\hat{\mathbb{P}}_\psi$) the von-Neumann entropy is zero, because the density operator is a
one-rank projector and its eigenvalues are $\lambda_1=1$ and the rest are zeros; thus, the
information contained in a pure state is maximal. The maximal uncertainty, in dimension four, is
represented by a maximally mixed states ($\tfrac{1}{4}\hat{\openone}_4$), with a value for the
von-Neumann entropy of $2$, in bits; because it is eigenvalues are all $\lambda_i=\tfrac{1}{4}$.
Then, in bipartites systems of qubits one has $0\leq S[\rho]\leq2$. Thus, the entropy for the GWLs
and Ws given in \eqref{eq:4} and \eqref{eq:0} will be bounded between these two values, being zero
when $p=1$ and $p=-1$ and maximum when $p=0$, respectively. The GWLs given in the equation
\eqref{eq:4}, has a simple eigenvalue given by $\tfrac{1+3p}{4}$, and three degenerates eigenvalues
with value $\frac{1-p}{4}$, this allows to obtain the von-Neumann entropy, given by
\begin{gather}
S_{AB}[\psi,p]=S[\rho_{\textrm{GWL}}]%
=-\tr{\rho_{\textrm{GWL}}\log_{2}\rho_{\textrm{GWL}}}
\nonumber\\
S_{AB}[\psi,p]=2-\tfrac{3(1-p)}{4}\log_{2} (1-p) - \tfrac{1+3p}{4}\log_{2} (1+3p)%
\nonumber\\
\label{eq:6}
S_{AB}[\psi,p]=2-\tfrac{1}{4}\log_2\left[\tfrac{(1+3p)^{3p+1}}{(1-p)^{3p-3}}\right].
\end{gather}
\par\noindent
This expression is independent of the values $z_i$ of GBps $\ket{\psi}$, in addition to being a
monotonic function of the mixing parameter $p$. Is clear from \eqref{eq:6} that the information
provided by the GWLs is minimal (maximum entropy) when $p=0$, while that the information is maximal
(minimum entropy) when $p=1$.
\\
\par\indent%
To determine the quantum information of each partition of the system
$\mathscr{H}_2\otimes\mathscr{H}_2$ contained in GWLs, given the equation \eqref{eq:4}, is
sufficient to take their partial traces, so that we have to
\begin{subequations}\label{eq:7}
\begin{eqnarray}
\label{eq:7a}
\hspace{-.5cm}\rho_{\textrm{GWL}}^A&=&\tr[B]{\rho_{\textrm{GWL}}}=%
\tfrac{1-p}{2}\hat{\openone}_2+p\widehat{\mathbb{W}}_\psi\widehat{\mathbb{W}}_\psi^\dag,
\\
\label{eq:7b}
\hspace{-.5cm}\rho_{\textrm{GWL}}^B&=&\tr[A]{\rho_{\textrm{GWL}}}=%
\tfrac{1-p}{2}\hat{\openone}_2+p(\widehat{\mathbb{W}}_\psi^T)(\widehat{\mathbb{W}}_\psi^T)^\dag.
\end{eqnarray}
\end{subequations}
Where $\widehat{\mathbb{W}}^{T}_\psi$ is the transposed matrix of $\widehat{\mathbb{W}}_\psi$. In
this context the transpose operation connects the density operator of both partitions and this
operation does not modify the eigenvalues of the reduced states. For this reason, the expressions
\eqref{eq:7a} and \eqref{eq:7b} show that the entropies of the reduced states are equal, so that
\begin{gather}
-\tr{\rho_{\textrm{GWL}}^A\log_2\rho_{\textrm{GWL}}^A}\equiv%
-\tr{\rho_{\textrm{GWL}}^B\log_2\rho_{\textrm{GWL}}^B},
\nonumber\\
 S_A(\psi,p)\equiv S_B(\psi,p).
\nonumber\\
\label{eq:10}
S_{A}(\psi,p)=S_{B}(\psi,p)=H_{2}\left(\tfrac{1+p\Delta_{\ket{\psi}}}{2}\right).
\end{gather}
Since $\widehat{\mathbb{W}}_\psi\widehat{\mathbb{W}}_\psi^\dag$ has two eigenvalues give by
$\tfrac{1}{2}\left(1\pm\Delta_{\ket{\psi}}\right)$. When $p=1$ in the equation \eqref{eq:10} one
has the EoF of pure state $\ket{\psi}$, given in the equation \eqref{eq:33}. On the other hand,
when $p=0$ the entropy of the reduced state is maximal, take the value of one bit, which
corresponds to a maximally mixed state in $\mathscr{H}_2$.
\\
\par\indent%
In order to quantify the conditional entropy, a projective measurement is required. We performed
this measurement on the partition $A$ of the bipartites system, in accordance with
\begin{subequations}\label{eq:8}
\begin{equation}\label{eq:8a}
\widehat{\Pi}_{m}^{A}=\widehat{\Pi}_{m}\otimes\hat{\openone}_{2}=%
\tfrac{1}{2}\left[\hat{\openone}_{2}+(-1)^{m}\hat{n}\cdot\vec{\sigma}\right]\otimes\hat{\openone}_{2},
\end{equation}
and projective measurement made on the partition $B$ is given by
\begin{equation}\label{eq:8b}
\widehat{\Pi}_{m}^{B}=\hat{\openone}_{2}\otimes\widehat{\Pi}_{m}=%
\hat{\openone}_{2}\otimes\tfrac{1}{2}\left[\hat{\openone}_{2}+(-1)^{m}\hat{n}\cdot\vec{\sigma}\right],
\end{equation}
\end{subequations}
with $m=0,1$. Here $\hat{n}=\sin(2\theta)\cos(\phi)\vi+\sin(2\theta)$$\sin(\phi)\vj+
\cos(2\theta)\vk$, is a unitary vector on the Bloch sphere, and
$\vec{\sigma}=\sigma_{x}\vi+\sigma_{y}\vj+\sigma_{z}\vk$ is the Pauli vector; while the set
$\{\vi,\vj,\vk\}$ is canonical basis of the Euclidean space $\mathbb{R}^3$. After this local
measurement on the density matrix $\rho_{GWL}$, given in the equation \eqref{eq:4}, the state of
the system becomes a hybrid quasi-classical state~\cite{Modi}, where both partitions have same
functional expression to a GWLs, in a two dimensional space. This is, using the L\"{u}der
rule~\cite{luders} for the partition $X$ we have that
\begin{equation}\label{eq:34}
\rho_{\textrm{GWL}}%
\xrightarrow{\hspace{.3cm}\widehat{\Pi}_{m}^{X}\hspace{.3cm}}%
\rho_{\textrm{GWL}|\Pi_{m}^{X}}=
\frac{(\widehat{\Pi}_{m}^{X})\rho_{\textrm{GWL}}(\widehat{\Pi}_{m}^{X})^{\dag}}{p_{m}^{X}},
\end{equation}
with $X=A,B$ and $p_{m}^{X}$ correspond to the probability of reaching that post-measurement states
in the partition $X$. This probability can be evaluated (see appendix~\ref{sec:apenA}) as
\begin{equation}\label{eq:15}
p_{m}^{X}=\braket{\widehat{\Pi}_{m}^{X}}_{\rho_{\textrm{GWL}}}=%
\tr{\widehat{\Pi}_{m}^{X}\rho_{\textrm{GWL}}}=%
\tfrac{1-p}{2}+p\braket{\widehat{\Pi}_m^X}_\psi.%
\end{equation}
Where $\braket{\widehat{\Pi}_m^X}_\psi$ is the transition probability of the GBps to the state
$\widehat{\Pi}_m^X$, which are evaluated for both partitions as
\begin{subequations}\label{eq:17}%
\begin{eqnarray}
\label{eq:17a}
\braket{\widehat{\Pi}_m^A}_\psi&=&\tr{\widehat{\mathbb{W}}_\psi^{\dag}\widehat{\Pi}_{m}\widehat{\mathbb{W}}_\psi},
\\
\label{eq:17b}
\braket{\widehat{\Pi}_m^B}_\psi&=&\tr{(\widehat{\mathbb{W}}_\psi^T)^{\dag}\widehat{\Pi}_{m}(\widehat{\mathbb{W}}_\psi^T)}.
\end{eqnarray}
\end{subequations}
Then, the mixing states obtained with the rule L\"uders \eqref{eq:34} for both partitions (see
appendix~\ref{sec:apenA}) are give by
\begin{subequations}\label{eq:14}
\begin{eqnarray}
\label{eq:14a}
\rho_{\textrm{GWL}|\Pi_{m}^{A}}&=&%
\widehat{\Pi}_{m}\otimes\left\{\tfrac{1-x_m(p)}{2}\hat{\openone}_2+x_m(p)\hat{\mathbb{P}}_B\right\},
\\
\label{eq:14b}
\rho_{\textrm{GWL}|\Pi_{m}^{B}}&=&%
\left\{\tfrac{1-y_m(p)}{2}\hat{\openone}_2+y_m(p)\hat{\mathbb{P}}_A\right\}\otimes\widehat{\Pi}_{m},
\end{eqnarray}
\end{subequations}
where $\hat{\mathbb{P}}_A\definido\ket{\widehat{\psi}_A}\bra{\widehat{\psi}_A}$ and
$\hat{\mathbb{P}}_B\definido\ket{\widehat{\psi}_B}\bra{\widehat{\psi}_B}$ are one-rank projectors,
defined by
\begin{subequations}\label{eq:19}
\begin{eqnarray}
\label{eq:19b}
\hspace{-.4cm}\hat{\mathbb{P}}_{A}&=&\sum_{i,j}%
\tfrac{\bra{i}(\widehat{\mathbb{W}}_\psi^T)^{\dag}\left(\hat{\openone}_{2}+(-1)^{m}\hat{n}\cdot\vec{\sigma}\right)
(\widehat{\mathbb{W}}_\psi^T)\ket{j}}{\tr{(\widehat{\mathbb{W}}_\psi^T)^{\dag}\left(\hat{\openone}_{2}+(-1)^{m}\hat{n}\cdot\vec{\sigma}\right)(\widehat{\mathbb{W}}_\psi^T)}}\,
\ket{j}\bra{i}.
\\
\label{eq:19a}
\hspace{-.4cm}\hat{\mathbb{P}}_{B}&=&\sum_{i,j}%
\tfrac{\bra{i}{\widehat{\mathbb{W}}}_\psi^{\dag}\left(\hat{\openone}_{2}+(-1)^{m}\hat{n}\cdot\vec{\sigma}\right)
\widehat{\mathbb{W}}_\psi\ket{j}}{\tr{{\widehat{\mathbb{W}}}_\psi^{\dag}\left(\hat{\openone}_{2}+(-1)^{m}\hat{n}\cdot\vec{\sigma}\right)\widehat{\mathbb{W}}_\psi}}\,
\ket{j}\bra{i},
\end{eqnarray}
\end{subequations}
The quantities $x_{m}(p)$ and $y_m(p)$ that appear in the equations \eqref{eq:14} are equivalents
to new mixing parameters of the GWLs in the partition $B$ and $A$, respectively. These are given by
(see appendix~\ref{sec:apenA})
\begin{subequations}\label{eq:22}%
\begin{eqnarray}
\label{eq:22a}
x_{m}(p)&=&\frac{p\langle\widehat{\Pi}_{m}^{A}\rangle_{\psi}}{\frac{1-p}{2}+p\langle\widehat{\Pi}_{m}^{A}\rangle_{\psi}}%
\\%
\label{eq:22b}
y_{m}(p)&=&\frac{p\langle\widehat{\Pi}_{m}^{B}\rangle_{\psi}}{\frac{1-p}{2}+p\langle\widehat{\Pi}_{m}^{B}\rangle_{\psi}}%
\end{eqnarray}
\end{subequations}
Noteworthy, that $x_{m}(p)$ and $y_m(p)$ are an injective functions of the mixing parameter $p$, so
$x_{m}(p)$ and $y_m(p)$ present the same variation range of $p$. It is important to see as well
that the two $x_{m}(p)$ or $y_m(p)$ are not independent, since the sum over all probabilities
($\sum_{m}\langle\widehat{\Pi}_{m}^{X}\rangle_{\psi}=1$ with $X=A,B$) impose a restriction on the
mixing parameters of the reduced states. This restrictions is given by
\begin{equation}\label{eq:18}
\sum_m\frac{x_m(p)}{1-x_m(p)}=\sum_m\frac{y_m(p)}{1-y_m(p)}=\frac{2p}{1-p}.
\end{equation}
Of the expressions indicated in \eqref{eq:7},  \eqref{eq:17}, \eqref{eq:14} and  \eqref{eq:19} it
is clear that the results of the meansurement process in the partition $A$ and $B$ are built with
the matrix $\widehat{\mathbb{W}}$ and $\widehat{\mathbb{W}}^T$, respectively. In this context, the
transpose operation gathered with exchange operator connects to the post-measurement mixing states
of both partitions.
\\
\par\indent%
Until now, a projective measurement on the partitions $A$ or $B$ projects the system into the
statistical ensembles $\left\{p_{m}^{A},\rho_{AB|\Pi_{m}^{A}}\right\}$ or
$\left\{p_{m}^{B},\rho_{AB|\Pi_{m}^{B}}\right\}$, respectively, quantifies the information in the
unmeasured partition by means of the quantum conditional entropy, given respectively by
\begin{subequations}\label{eq:44}%
\begin{eqnarray}
\label{eq:44b}
\hspace{-.5cm}S_{A|\{\Pi_m^B\}}\left(\rho_{AB}\right)%
\hspace{-.2cm}&=&\hspace{-.2cm}\min_{\{\Pi_m^B\}}\sum_{m}\braket{\Pi_m^B}_\rho \
S_{A|\Pi_m^B}\left(\rho_{AB}\right),
\\%
\label{eq:44a}
\hspace{-.5cm}S_{B|\{\Pi_m^A\}}\left(\rho_{AB}\right)%
\hspace{-.2cm}&=&\hspace{-.2cm}\min_{\{\Pi_m^A\}}\sum_{m}\braket{\Pi_m^A}_\rho \
S_{B|\Pi_m^A}\left(\rho_{AB}\right),
\end{eqnarray}
\end{subequations}
where $S_{A|\Pi_m^B}\left(\rho_{AB}\right)$ and $S_{B|\Pi_m^A}\left(\rho_{AB}\right)$ are the
von-Neumann entropy of the partition $A$ and $B$ of $\rho_{AB}$ obtained after the projective
measurements ${\Pi_m^B}$ or ${\Pi_m^A}$, respectively. The entropy might give different results
depending on the basis choice, a minimization is taken over all possible one-rank measurements so
that minimization chooses the measurement of a partition that extracts as much information as
possible of the other partition. The entropy after of the measurement in the partition $A$ it is
given by
\begin{equation}\label{eq:20}
\begin{split}
S_{B|\{\Pi_m^A\}}(\psi,p)%
&=\min_{\{\Pi_m^A\}}\sum_{m}p_m^AS_{B|\Pi_m^A}(\psi,p),\\
=\tfrac{1}{2}\min_{\{\Pi_m^A\}}&\sum_{m}\tfrac{1-p}{1-x_m(p)} H_2\left(\tfrac{1+x_m(p)}{2}\right).
\end{split}
\end{equation}
Here the probability $p_m^A$ is replaced by the expression \eqref{eq:15}, while the probability
$\braket{\widehat{\Pi}_m^A}_\psi$ is written in terms of the mixing parameter $x_m(p)$ using
\eqref{eq:22a}. The hard step in the evaluation of the quantum conditional entropy is usually the
optimization of the conditional entropy $S_{B|\Pi_{m}^{A}}$ over all projective measurements.
However, in the Appendix~\ref{sec:apenB} we showed that the process of minimizing for conditional
entropy consists in finding the values of $x_{m}(p)$ that minimize the probability
$\braket{\widehat{\Pi}^A_m}_\psi$. Such that the conditional entropy of the partition $B$ have the
form
\begin{equation}\label{eq:21}
S_{B|\{\Pi_m^A\}}(\psi,p)=F_p(\underline{x}_{0})+F_p(\underline{x}_{1}),
\end{equation}
where
\begin{equation}\label{eq:23}
F_p(x)=\tfrac{1-p}{2(1-x)}H_2\left(\tfrac{1+x}{2}\right),
\end{equation}
while the values of $\underline{x}_{0}$ and $\underline{x}_{1}$ minimize and maximize the
probability $\braket{\widehat{\Pi}_m^A}_\psi$ in the equation \eqref{eq:22a}, respectively (see
appendix~\ref{sec:apenB}). Namely, $\underline{x}_{0}$ is obtained when the probability
$\braket{\widehat{\Pi}^A_m}_\psi$ is minimized,
\begin{subequations}\label{eq:24}%
\begin{equation}\label{eq:24a}
\underline{x}_{0}=
\frac{p\braket{\widehat{\Pi}_{0}^{A}}^{\min}_{\psi}}{\frac{1-p}{2}+p\braket{\widehat{\Pi}_{0}^{A}}^{\min}_{\psi}}\,,%
\end{equation}
but $\underline{x}_{1}$ is obtained from to relation \eqref{eq:18}, finding that
\begin{equation}\label{eq:24b}
\begin{split}
\underline{x}_{1}&=%
\frac{2p-(1+p)\underline{x}_0}{1+p-2\underline{x}_0}=
\frac{p(1-\braket{\widehat{\Pi}_{0}^{A}}^{\min}_\psi)}{\frac{1-p}{2}+p(1-\braket{\widehat{\Pi}_{0}^{A}}^{\min}_\psi)}
\\
&=\frac{p\braket{\widehat{\Pi}_{1}^{A}}^{\max}_{\psi}}{\frac{1-p}{2}+p\braket{\widehat{\Pi}_{1}^{A}}^{\max}_{\psi}}.
\end{split}
\end{equation}
\end{subequations}
The probability $\braket{\widehat{\Pi}_0^A}_\psi$ presents oscillations around $1/2$ with
ammplitude $A_\psi$ (see appendix \ref{sec:apenB}), thereby the value of
$\braket{\widehat{\Pi}_0^A}_\psi^{\min}=\tfrac{1}{2}-A_\psi$, and $\underline{x}_{0}$ and
$\underline{x}_{1}$ can be written as
\begin{equation}\label{eq:57}
\underline{x}_{0}=\frac{p(1-2A_\psi)}{1-2pA_\psi},%
\quad%
\underline{x}_{1}=\frac{p(1+2A_\psi)}{1+2pA_\psi}.
\end{equation}
A straightforward calculation show that
\begin{equation}\label{eq:25}
A_\psi=\frac{1}{2}\sqrt{\sum_{i=1}^{3}\left(\tr{\widehat{\mathbb{W}}^{\dag}_\psi\sigma_{i}\widehat{\mathbb{W}}_\psi}\right)^2}=%
\tfrac{1}{2}\Delta_{\ket{\psi}}.
\end{equation}
The equation \eqref{eq:21} is an exact analytical expression for the conditional entropy after a
measurement in partition $A$. The aforementioned procedure can be applied to obtain the conditional
entropy $S_{A|\{\Pi_m^B\}}(\psi,p)$, after a measurement in partition $B$. The same result is
obtained, except that instead of the matrix $\mathbb{\mathbb{W}}_\psi$, its transpose is used. In
addition, the mixing parameter $x_m(p)$ must be replaced by $y_m(p)$, namely,
\begin{gather}
S_{A|\{\Pi_m^B\}}(\psi,p)=\min_{\{\Pi_m^B\}}\sum_{m}p_m^BS_{A|\Pi_m^B}(\psi,p),
\nonumber%
\intertext{giving }%
S_{A|\{\Pi_m^B\}}(\psi,p)=\tfrac{1}{2}\min_{\{\Pi_m^B\}}\sum_{m}\tfrac{1-p}{1-y_m(p)}
H_2\left(\tfrac{1+y_m(p)}{2}\right).
\nonumber%
\intertext{This implies that the minimized condition entropy take the form (see
appendix~\ref{sec:apenB})}
\label{eq:26}
S_{A|\{\Pi_m^B\}}(\psi,p)=F_p(\underline{y}_{0})+F_p(\underline{y}_{1}),
\end{gather}
with
\begin{equation}\label{eq:28}
\underline{y}_{0}=\frac{p(1-2B_\psi)}{1-2pB_\psi}%
\quad\textrm{and}\quad%
\underline{y}_{1}=\frac{p(1+2B_\psi)}{1+2pB_\psi}\,,
\end{equation}
futhermore
\begin{equation}\label{eq:29}
B_\psi=\frac{1}{2}\sqrt{\sum_{i=1}^{3}\left(\tr{(\widehat{\mathbb{W}}_\psi^T)^{\dag}\sigma_{i}(\widehat{\mathbb{W}}_\psi^T)}\right)^2}=%
\tfrac{1}{2}\Delta_{\ket{\psi}}.%
\end{equation}
It is important to indicate that the value of $B_\psi$ is coincident with the value of $A_\psi$,
and both quantity are monotonous functions of the concurrence of the GBps $\ket{\psi}$. For this
reason the conditional entropy of both partitions are the same, and as well this amounts are
monotonous functions of the concurrence of the GBps.
\\
\par\indent%
Finally, the QD or LII is defined as the difference between the total correlation (or mutual
information) and classical correlations (or conditional mutual information) coded in the same
state. The quantum mutual informations or total correlation is a measure of how much information
grows in a bipartite system when partitions are observed together. This quantity is defined as
$\mathcal{I}_{AB}=S[\rho_A]+S[\rho_B]-S[\rho_{AB}]$. The classical correlations or conditional
mutual information measured in the partition $A$ and $B$ are written as
$\mathcal{J}_{\overrightarrow{AB}}=S[\rho_B]-S_{B|\{\Pi^{A}_m\}}(\rho_{AB})$ and
$\mathcal{J}_{\overleftarrow{AB}}=S[\rho_B]-S_{A|\{\Pi^{B}_m\}}(\rho_{AB})$, respectively. These
quantities measure the gain of information in the partition when the other is measured. Then the QD
or LII of any state $\rho_{AB}$, when performing measured on the partition $A$, can be written as
\begin{subequations}\label{eq:43}%
\begin{eqnarray}
\hspace{-.4cm}\delta_{\overrightarrow{AB}}(\rho_{AB})&=&\mathcal{I}_{AB}(\rho_{AB})-\mathcal{J}_{\overrightarrow{AB}}(\rho_{AB})%
\nonumber\\%
\label{eq:43a}
&=&S[\rho_{A}]-S[\rho_{AB}]+S_{B|\{\Pi_m^A\}}(\rho_{AB}),
\end{eqnarray}
and when performing measured on the partition $B$ it is given by
\begin{eqnarray}
\hspace{-.4cm}\delta_{\overleftarrow{AB}}(\rho_{AB})&=&%
\mathcal{I}_{AB}(\rho_{AB})-\mathcal{J}_{\overleftarrow{AB}}(\rho_{AB})%
\nonumber\\%
\label{eq:43b}
&=&S[\rho_{B}]-S[\rho_{AB}]+S_{A|\{\Pi_m^B\}}(\rho_{AB}).
\end{eqnarray}
\end{subequations}
Generally the QD of mixing states is asymmetric, i.e.
$\delta_{\overrightarrow{AB}}\neq\delta_{\overleftarrow{AB}}$. This allow to study the average of
LII, defined as $\varpi_{A|B}^{+}=(\delta_{\overrightarrow{AB}}+\delta_{\overleftarrow{AB}})/2$,
and the balance of LII, defined as
$\varpi_{A|B}^{-}=(\delta_{\overrightarrow{AB}}-\delta_{\overleftarrow{AB}})/2$ (see
reference~\cite{Fanchinia}). Nevertheless, a straightforward calculation showed that \eqref{eq:25}
and \eqref{eq:29} are coincident (see appendix~\ref{sec:apenB}), being iquals the QD of the GWLs in
both partitions, therefore the balance of LII is zero and the average of LII is same that the QD
for GWLs. If we take the explicit forms of the entropies given in the equation \eqref{eq:43} we can
obtain the exact analytical expressions for the GWLs, being
\begin{eqnarray}
\delta_{AB}(\psi,p)&=&%
\delta_{\overleftarrow{AB}}(\psi,p)=\delta_{\overrightarrow{AB}}(\psi,p)%
\nonumber\\%
&=&-2+\tfrac{1}{4}\log_2\left[\tfrac{(1+3p)^{3p+1}}{(1-p)^{3p-3}}\right]+%
 H_{2}\left(\tfrac{1+p\Delta_{\psi}}{2}\right)
\nonumber\\%
\label{eq:30}
&&+\tfrac{1-p\Delta_{\ket{\psi}}}{2}H_2\left(\tfrac{1+p(1-2\Delta_{\ket{\psi}})}{2(1-p\Delta_{\ket{\psi}})}\right)
\\%
&&+\tfrac{1+p\Delta_{\ket{\psi}}}{2}H_2\left(\tfrac{1+p(1+2\Delta_{\ket{\psi}})}{2(1+p\Delta_{\ket{\psi}})}\right)
\nonumber%
\end{eqnarray}
The QD is zero when $p=0$ in the expression \eqref{eq:30}, and the QD is coincident with the EoF of
GWps when $p=1$. In these cases it is take into account that $H_2(\tfrac{1}{2})=1$ and $H_2(1)=0$.
Then, the QD is symmetrical and also is a monotonous function of the concurrence $C[\ket{\psi}]$ of
the GBps. So, all the GBps with the same concurrence have equal QD, forming equivalence classes.

\section{Examples}
\label{sec:examples}%
In order to illustrate the behavior of the QD of the GWLs, and its dependence with the mixing
parametrer $p$ and the concurrence of GBps, we consider the four pure states shown below
\vspace{-.3cm}%
\begin{subequations}\label{eq:1}
\begin{align}
\label{eq:1a}
&\hspace{-.2cm}\ket{\psi_1}=\tfrac{\sqrt{7}}{8}\ket{00}+\tfrac{3\sqrt{5}}{8}\ket{01}+\tfrac{\sqrt{5}}{8}\ket{10}+\tfrac{\sqrt{7}}{8}\ket{11},
\\
\label{eq:1b}
&\hspace{-.2cm}\ket{\psi_2}=-\tfrac{1}{2}\ket{00}-\tfrac{\sqrt{2}}{2}\ket{01}+\tfrac{\sqrt{2}}{3}\ket{10}+\tfrac{1}{6}\ket{11},
\\
\label{eq:1c}
&\hspace{-.2cm}\ket{\psi_3}={\scriptstyle\sqrt{\tfrac{9}{40}}}\ket{00}+{\scriptstyle\sqrt{\tfrac{3}{20}}}\ket{01}+{\scriptstyle\sqrt{\tfrac{3}{5}}}\ket{10}-{\scriptstyle\sqrt{\tfrac{1}{40}}}\ket{11},
\\
\label{eq:1d}
&\hspace{-.2cm}\ket{\Psi_{+}}=\tfrac{\sqrt{2}}{2}\ket{00}+\tfrac{\sqrt{2}}{2}\ket{11}.
\end{align}
\end{subequations}
These pure states can be represented in terms of the matrix $\widehat{\mathbb{W}}_\psi$, given in
the equation \eqref{eq:2}, according to
\begin{subequations}\label{eq:31}%
\begin{align}
\label{eq:31a}
\mathbb{W}_{\psi_1}&=\begin{bmatrix} \tfrac{\sqrt{7}}{8} & \tfrac{3\sqrt{5}}{8}\\*[.1cm] \tfrac{\sqrt{5}}{8} & \tfrac{\sqrt{7}}{8}\\\end{bmatrix},%
&%
\mathbb{W}_{\psi_2}&=\begin{bmatrix} -\tfrac{1}{2} & -\tfrac{\sqrt{2}}{2}\\*[.1cm] \tfrac{\sqrt{2}}{3} & \tfrac{1}{6}\\\end{bmatrix},%
\\*[0cm]%
\label{eq:31b}
\mathbb{W}_{\psi_3}&=\begin{bmatrix} \sqrt{\tfrac{9}{40}} & \sqrt{\tfrac{3}{20}}\\*[.1cm] \sqrt{\tfrac{3}{5}} & -\sqrt{\tfrac{1}{40}}\\\end{bmatrix},%
&%
\mathbb{W}_{\Psi_{+}}&=\begin{bmatrix} \tfrac{\sqrt{2}}{2} & 0\\*[.1cm] 0 & \tfrac{\sqrt{2}}{2}\\\end{bmatrix}.%
\end{align}
\end{subequations}
The concurrences of these pure states are given by
$\left\{\tfrac{1}{4},\tfrac{1}{2},\tfrac{3}{4},1\right\}$, respectively, which are determined using
the equation \eqref{eq:34}. All those GWLs built whit any pure state that presents these
concurrency values will have the same QD as well as the same EoF. On the other hand, the density
matrix for the GWLs built with the pure states given in the equations \eqref{eq:1a}, \eqref{eq:1b}
and \eqref{eq:1c} are non--X states, while the pure state given in the equation \eqref{eq:1d} is a
X--state. Thus, using the expression \eqref{eq:4} we obtained that the density matrices for the
GWLs built with the aforementioned pure states have the following form
\begin{widetext}
\begin{align*}
\rho_{GWL}(\psi_1,p)&=\tfrac{1}{64}%
\begin{bmatrix}
16-9p       & 3\sqrt{35}p & \sqrt{35}p & 7p \\
3\sqrt{35}p & 16+29p      & 15p        & 3\sqrt{35}p \\
\sqrt{35}p  & 15p         & 16-11p     & \sqrt{35}p \\
7p          & 3\sqrt{35}p & \sqrt{35}p & 16-9p \\
\end{bmatrix},
&\hspace{0cm}%
\rho_{GWL}(\psi_2,p)&=\tfrac{1}{36}%
\begin{bmatrix}
 9          & 9\sqrt{2}p & -6\sqrt{2}p & -3p \\
 9\sqrt{2}p & 9(1+p)     & -12p        & -3\sqrt{2}p \\
-6\sqrt{2}p &-12p        & 9-p         & 2\sqrt{2}p \\
-3p         &-3\sqrt{2}p & 2\sqrt{2}p  & 9-8p \\
\end{bmatrix},
\\
\rho_{GWL}(\psi_3,p)&=\tfrac{1}{40}%
\begin{bmatrix}
 10-p       & 3\sqrt{6}p & 6\sqrt{6}p & -3p        \\
 3\sqrt{6}p & 10-4p      & 12p        &-\sqrt{6}p  \\
 6\sqrt{6}p & 12p        & 2(5+7p)    &-2\sqrt{6}p \\
-3p         &-\sqrt{6}p  &-2\sqrt{6}p & 10-9p      \\
\end{bmatrix},
&\hspace{0cm}%
\rho_{GWL}(\Psi_+,p)&=\tfrac{1}{4}%
\begin{bmatrix}
 1+p& 0  & 0  & 2p \\
 0  & 1-p& 0  & 0  \\
 0  & 0  & 1-p& 0  \\
 2p & 0  & 0  & 1+p\\
\end{bmatrix},
\end{align*}
\end{widetext}
For that a GWLs to present the form of a X--state the matrix $\widehat{\mathbb{W}}_\psi$ of the
pure state $\ket{\psi}$ should have the form of one diagonal or antidiagonal matrix; any other way,
the GWLs are non--X states. However, the EoF and QD of GWLs only depend of the concurrence of the
GBps and it does not depend on the topology that the mixing states possesses. For example, the GWLs
built with the pure state
\begin{displaymath}
\ket{\psi_5}=-\tfrac{\sqrt{2+\sqrt{3}}}{2}\ket{01}+\tfrac{\sqrt{2-\sqrt{3}}}{2}\ket{10}
\end{displaymath}
have a density matrix in form of non--X state but their QD, as well as the EoF, are the same
obtained with the pure state $\ket{\psi_2}$ given in \eqref{eq:1b}.
\\
\begin{figure}[t]
\begin{picture}(0,4.5)(3.5,0)
\put(-.8,0){%
\put(0,0){\includegraphics[height=4.5cm,width=8cm]{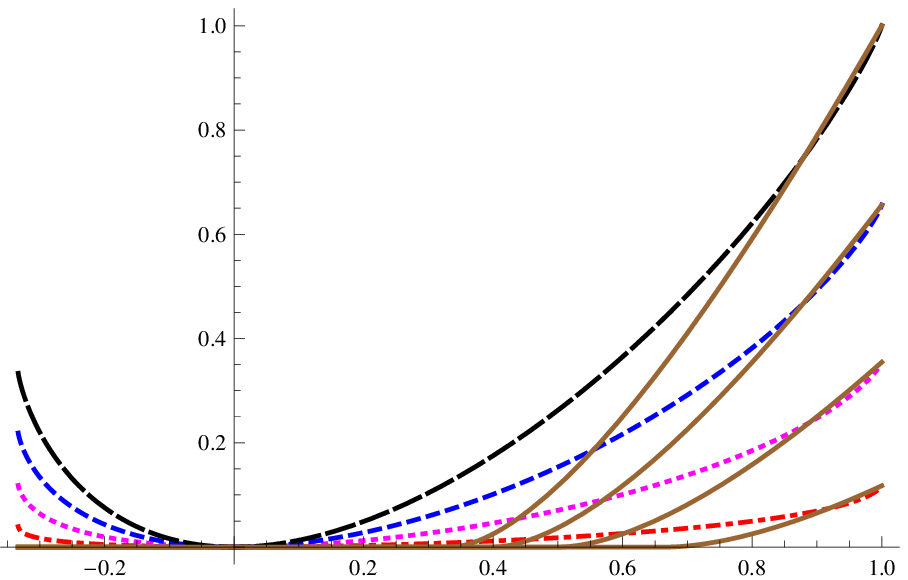}}%
\put(8,.2){\footnotesize$p$}\put(1.5,4.6){\footnotesize EoF, $\delta_{AB}$}%
\put(7.9,.5){%
\put(0,3.5){\footnotesize$C_{\max}$}%
\put(0,2.2){\footnotesize$C_{3}$}%
\put(0,1){\footnotesize$C_{2}$}%
\put(0,.1){\footnotesize$C_{1}$}%
}%
\put(2.4,3.8){
 \put(.8,0.3){\footnotesize\underline{States}}
 \psline[linecolor=brown](0,0)(.5,0)%
 \put(.6,-.1){\footnotesize EoF for GWLs}%
 \psline[linecolor=black](0,-.4)(.5,-.4)%
 \put(.6,-.5){\footnotesize QD for GWLs $\ket{\Psi_{+}}$}%
 \multiput(0,-.8)(.15,0){4}{\psline[linecolor=blue](0,0)(.1,0)}%
 \put(.6,-.9){\footnotesize QD for GWLs $\ket{\psi_3}$}%
 \multiput(0,-1.2)(.15,0){4}{\psline[linecolor=magenta](0,0)(.07,0)}%
 \put(.6,-1.3){\footnotesize QD for GWLs $\ket{\psi_2}$}%
 \psline[linecolor=red](0,-1.6)(.07,-1.6)\psline[linecolor=red](.15,-1.6)(.35,-1.6)\psline[linecolor=red](.43,-1.6)(.5,-1.6)%
 \put(.6,-1.7){\footnotesize QD for GWLs $\ket{\psi_1}$}%
           }%
}%
\end{picture}
\caption{(color online) Plot of the EoF (line solid brown) and QD, in function of mixing parameter
$p$, for GWLs with a discrete pure state $\ket{\Psi_{+}}$, $\ket{\psi_3}$ (line blue dashed),
$\ket{\psi_2}$ (magenta dotted line) and $\ket{\psi_4}$ (dot-dashed red line) those concurrences
are equal to $C_{\max}=1$, $C_3=\tfrac{3}{4}$, $C_2=\tfrac{1}{2}$ and $C_1=\tfrac{1}{4}$,
respectively.} \label{fig:discretestates}
\end{figure}
\par
\begin{figure}[t]
\begin{picture}(0,4.5)(3.5,0)
\put(-.8,0){%
\put(0,0){\includegraphics[height=4.5cm,width=8cm]{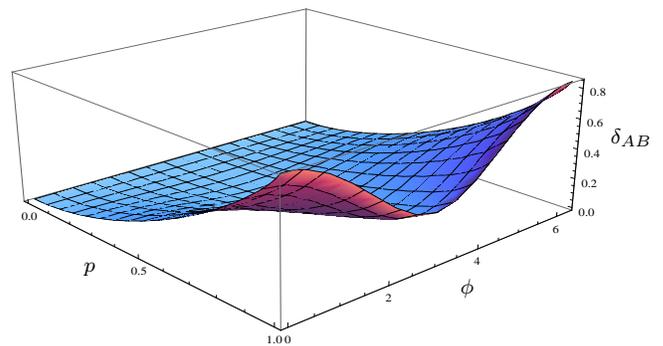}}%
\put(1,1){\footnotesize$p$}%
\put(6,.7){\footnotesize$\phi$}%
\put(8,2.7){\footnotesize$\delta_{AB}$}%

}%
\end{picture}
\caption[color online]{Plot of the QD for GWLs in function of mixing parameter $p$ and the local
angle of phase, present in the pure state $\ket{\Psi_6}$.} \label{fig:QDconFaseLocal}
\end{figure}

\indent%
The QD for the states indicated in the equations \eqref{eq:1} are sketched in the
Fig.~\ref{fig:discretestates}, together with the EoF calculated by the equation \eqref{eq:33}.
Finally, let us study the effect of incorporating a local phase into the pure state with which the
GWLs are built, for this we consider the following state
\begin{equation}\label{eq:12}
\begin{split}
\ket{\psi_6}=&{-\tfrac{\sqrt{2}}{6}\exp^{i\phi_1}}\ket{00}+%
             {\tfrac{\sqrt{2}}{3}\exp^{i\phi_2}}\ket{01}+%
             \\
             &+{\tfrac{\sqrt{2}}{2}\exp^{i\phi_3}}\ket{10}+%
             {\tfrac{\sqrt{2}}{3}\exp^{i\phi_4}}\ket{11},
\end{split}
\end{equation}
its concurrence can be written as
\begin{equation*}%
C[\ket{\psi_6}]%
=\tfrac{2}{9}\left|3\exp^{i(\phi_2+\phi_3)}+\exp^{i(\phi_1+\phi_4)}\right|
\end{equation*}
\begin{equation}\label{eq:13}
\begin{split}
C[\ket{\psi_6}]%
&=\tfrac{2}{9}\left|3+\exp^{i\phi}\right|=%
\tfrac{2}{9}\left|1+3\exp^{-i\phi}\right|%
\\
&=\tfrac{2}{9}\sqrt{10+6\cos\phi}.
\end{split}
\end{equation}
where $\phi\definido\phi_1+\phi_4-(\phi_2+\phi_3)$. The value of the concurrence given in the
equation \eqref{eq:13} does not change if $\phi$ is changed by $-\phi$, so that the QD and EoF are
same if the phase take any position in the component of pure state $\ket{\psi_5}$. Now the QD is
function of mixing parameter and angle of phase $\phi$, whose graph is shown in the
Fig.~\ref{fig:QDconFaseLocal}.

\section{Results}
\label{sec:Result}%
In Fig.~\ref{fig:discretestates} it can be observed that the QD is a monotonous function that grows
with the increase of the concurrence of GBps, but the QD of these states is not a monotonous
function of its own EoF for all values of the mixing parameter. The EoF and QD are coincident only
in three values $p=0$, $p=p_i$ and $p=1$, where $p_i$ is the value of the mixing parameter for
which the EoF intercepts with the QD, it is found numerically (since the equation that determines
the value of $p_i$ is transcendental) and are reported in Table~\ref{tab:1}. When $p>p_i$ we have
that $\textrm{EoF}(\psi,p)>\delta_{AB}(\psi,p)$ while that
$\textrm{EoF}(\psi,p)<\delta_{AB}(\psi,p)$ when $p<p_i$. We note that in the interval $p>p_i$ the
QD and EoF are very close to each other, but they present more discrepancies in the interval
$p_c<p<p_i$, beging $p_c$ the critical value of the mixing parameter that limits the border between
entanglement and separability of GWLs, which is obtained in the equation \eqref{eq:3}. In
Fig.~\ref{fig:discretestates} we observed that the GWLs is entanglement when $p>pc$. The values of
$p_c$ for the GWLs, built with the GBsp given in \eqref{eq:1}, are showed in Table~\ref{tab:1}. In
the interval $-\tfrac{1}{3}\le p\le p_i$ the GWLs contain mixing states that maintain a correlation
between the partitions of system, which is not associated with entanglement; in this sense it is
said that QD presents quantum correlations that go beyond entanglement. Also we observed that the
maximum value of QD and EoF is reached for those states that have maximum value of concurrence, so
that the QWs, WPs and BWLs they have more QD and Entanglement.
\\
\par\indent%
The QD and EoF of the GWLs is invariant under local unitary transformation, but these can reduce or
increase before a general unitary transformation, changing the correlations of the DWs.

\begin{table}[h]
\caption{Critical values $p_c$, the intersection point $p_i$ and the mixing parameter $p_b$ where
Bell inequality is violated}\label{tab:1}
\begin{ruledtabular}
\begin{tabular}{ccccc}
System     & $\ket{\psi_1}$ & $\ket{\psi_2}$ & $\ket{\psi_3}$ & $\ket{\Psi_{+}}$
\\%
Value $p_c$& $2/3$          & $1/2$          & $2/5$     & $1/3$
\\%
Value $p_i$& $0.919$        & $0.888$        & $0.878$   & $0.879$
\end{tabular}
\end{ruledtabular}
\end{table}
%

\section{Conclusion}
\label{sec:Conclusion}%
An exact analytical solution of QD for Generalized Werner-Like non--X states
have bee found. The optimization process involved in minimizing the conditional
entropy is solved in an analitical form. The QD obtained is symmetric and
increases with the concurrence of the GBps with which is built the GWLs. The
maximum value obtained for QD is only for WPs, QWs and BWLs. The Ws and GWLs
present diferent entanglement and QD, and only are coincident when the
transformation \eqref{eq:53} applies.

\begin{acknowledgments}
We are thankful to Prof.~Manuel Rodriguez and PhD. Naryttza Diaz-Fortier 
for his fruitful discussions, comments and remarks on the final version of this article. We are
particularly grateful for the comments of Dr. Y. Huang (Department of Physics, University of
California, Berkeley).
\end{acknowledgments}

\appendix%
\section{Projective measurement onto pure state and GWLs}%
\label{sec:apenA}%
Let $\mathbb{U}=[U_{ij}]$ be unitary transformation, the bases
$\{\ket{\pi_m}\}$ is unitarily equivalent to the computational bases
$\{\ket{i}\}$ if $\ket{\pi_m}=\sum_iU_{mi}\ket{i}$. The projector, associated
to these measurement are
\begin{equation}\label{eq:apenA1}%
\widehat{\Pi}_m=\ket{\pi_m}\bra{\pi_m}=%
\sum_{ij}U_{im}\overline{U}_{jm}\ket{i}\bra{j},
\end{equation}
where $\overline{U}_{jm}$ is the complex conjugate of $U_{jm}$. The projectors
associated to local projective measurement in the partition $A$ of a bipartite
system are
\begin{equation}\label{eq:apenA2}%
\widehat{\Pi}_m^A=\widehat{\Pi}_m\otimes\hat{\openone}=%
\sum_{ijk}U_{im}\overline{U}_{jm}\ket{ik}\bra{jk},
\end{equation}
where the identity operator $\hat{\openone}$ has been replaced by the sum of projectors
$\sum_k\ket{k}\bra{k}$. On the other hand, any pure state $\ket{\psi}$ that belongs to
$\mathscr{H}\otimes\mathscr{H}$ can be written in terms of computational basis as
\begin{equation}\label{eq:apenA3}%
\ket{\psi}=\sum_{ij}\psi_{ij}\ket{ij}%
\quad\textrm{with}\quad \sum_{ij}\psi_{ij}\overline{\psi}_{ij}=1.
\end{equation}
In order to simplify our results we define the matrix
$\widehat{\mathbb{W}}_\psi$, whose elements are $\psi_{ij}$, so the
normalization condition can be written as
\begin{gather}
\sum_{ij}\psi_{ij}\overline{\psi}_{ij}=1\;\Longrightarrow\;
\sum_{ij}[\widehat{\mathbb{W}}_\psi]_{i\times j}[\overline{\widehat{\mathbb{W}}_\psi}]_{i\times j}=1%
\nonumber\\
\sum_{ij}[\widehat{\mathbb{W}}_\psi]_{i\times
j}[\widehat{\mathbb{W}}_\psi^\dag]_{j\times i}=1
\;\Longrightarrow\;%
\sum_{i}[\widehat{\mathbb{W}}_\psi\widehat{\mathbb{W}}_\psi^\dag]_{i\times i}=1%
\nonumber\\
\label{eq:apenA4}%
\therefore\;
\tr{\widehat{\mathbb{W}}_\psi\widehat{\mathbb{W}}_\psi^\dag}=\tr{\widehat{\mathbb{W}}_\psi^\dag\widehat{\mathbb{W}}_\psi}=1
\end{gather}
The representation of a pure state in terms of the density matrix is given by
the following rank-one projector,
\begin{equation}\label{eq:apenA6}%
\ket{\psi}\bra{\psi}=\sum_{ijk\ell}\psi_{ij}\overline{\psi}_{k\ell}\ket{ij}\bra{k\ell},
\end{equation}
the reduced states are obtained by taking partial trace over both partitions,
thus, for partition $A$ we have that
\begin{gather}
\rho_A(\psi)=\tr[B]{\ket{\psi}\bra{\psi}}=
\sum_{ijk\ell}\psi_{ij}\overline{\psi}_{k\ell}\tr[B]{\ket{ij}\bra{k\ell}}%
\nonumber\\%
=\sum_{ijk\ell}\psi_{ij}\overline{\psi}_{k\ell}\delta_{j\ell}\ket{i}\bra{k}
=\sum_{ijk}\psi_{ij}\overline{\psi}_{kj}\ket{i}\bra{k}.
\nonumber\\%
=\sum_{ijk}[\widehat{\mathbb{W}}_\psi]_{i\times j}[\widehat{\mathbb{W}}_\psi^\dag]_{j\times k}\ket{i}\bra{k}%
=\sum_{ik}[\widehat{\mathbb{W}}_\psi\widehat{\mathbb{W}}_\psi^\dag]_{i\times k}\ket{i}\bra{k}.%
\nonumber\\%
\label{eq:apenA7}%
\therefore\quad\rho_A(\psi)=\widehat{\mathbb{W}}_\psi\widehat{\mathbb{W}}_\psi^\dag
\end{gather}
and for partition $B$ we have,
\begin{gather}
\rho_B(\psi)=\tr[A]{\ket{\psi}\bra{\psi}}=
\sum_{ijk\ell}\psi_{ij}\overline{\psi}_{k\ell}\tr[A]{\ket{ij}\bra{k\ell}}%
\nonumber\\%
=\sum_{ijk\ell}\psi_{ij}\overline{\psi}_{k\ell}\delta_{ik}\ket{j}\bra{\ell}
=\sum_{ij\ell}\psi_{ij}\overline{\psi}_{i\ell}\ket{j}\bra{\ell}.
\nonumber\\%
=\sum_{ij\ell}[\widehat{\mathbb{W}}_\psi]_{i\times j}[\widehat{\mathbb{W}}_\psi^\dag]_{\ell\times i}\ket{j}\bra{\ell}%
=\sum_{ij\ell}[\widehat{\mathbb{W}}_\psi^T]_{j\times i}[(\widehat{\mathbb{W}}_\psi^T)^\dag]_{i\times\ell}\ket{j}\bra{\ell}%
\nonumber\\%
=\sum_{j\ell}[(\widehat{\mathbb{W}}_\psi^T)(\widehat{\mathbb{W}}_\psi^T)^\dag]_{j\times \ell}\ket{j}\bra{\ell}.%
\nonumber\\%
\label{eq:apenA8}%
\quad\therefore\quad\rho_B(\psi)=(\widehat{\mathbb{W}}_\psi^T)(\widehat{\mathbb{W}}_\psi^T)^\dag
\end{gather}
This shows that partition $B$ can be accessed through the transpose operation.
On the other hand, the probability of obtaining a result after the local
projective measurement \eqref{eq:apenA2} when the system is initially in the
pure state $\ket{\psi}$ is given by
\begin{equation}\label{eq:apenA5}%
\braket{\widehat{\Pi}_m^A}_\psi=\bra{\psi}\widehat{\Pi}_m^A\ket{\psi}=%
\sum_{ijk}\overline{\psi}_{ik}\psi_{jk}U_{im}\overline{U}_{jm},
\end{equation}
in which equation \eqref{eq:apenA2} has been used. The last expression can be
written in matrix form as,
\begin{gather}
\braket{\widehat{\Pi}_m^A}_\psi=\sum_{ijk}\overline{\psi}_{ik}U_{im}\overline{U}_{jm}\psi_{jk}%
\nonumber\\
\sum_{ijk}
[\widehat{\mathbb{W}}_\psi^\dag]_{k\times i}[\widehat{\Pi}_m]_{i\times j}[\widehat{\mathbb{W}}_\psi]_{j\times k}%
=\tr{\widehat{\mathbb{W}}_\psi^\dag\widehat{\Pi}_m\widehat{\mathbb{W}}_\psi}
\nonumber\\
\label{eq:apenA9}%
\braket{\widehat{\Pi}_m^A}_\psi=
\tr{\widehat{\mathbb{W}}_\psi\widehat{\mathbb{W}}_\psi^\dag\widehat{\Pi}_m}%
\equiv\braket{\widehat{\Pi}_m}_{\rho_A(\psi)},
\end{gather}
where we have used the expressions \eqref{eq:apenA1} and \eqref{eq:apenA7}. If
the measurement is performed on partition $B$ then
\begin{equation}\label{eq:apenA10}%
\braket{\widehat{\Pi}_m^B}_\psi=\braket{\widehat{\Pi}_m}_{\rho_B(\psi)}=%
\tr{(\widehat{\mathbb{W}}_\psi^T)(\widehat{\mathbb{W}}_\psi^T)^\dag\widehat{\Pi}_m}.%
\end{equation}
Now, we perform a projective measurement on partition $A$ of the pure state
$\ket{\psi}$. In order to do this, we apply L\"uders rule~\cite{luders} to pure
state $\ket{\psi}$, obtaining
\begin{gather}
\label{eq:apenA15}
\ket{\psi}\bra{\psi}\Big|_{\Pi_m^A}=%
\frac{\widehat{\Pi}_m^A\ket{\psi}\bra{\psi}(\widehat{\Pi}_m^A)^\dag}{\braket{\widehat{\Pi}_m^A}_\psi},%
\\%
=\tfrac{1}{\braket{\widehat{\Pi}_m^A}_\psi}\sum_{ijk}\sum_{rst}%
\psi_{jk}\overline{\psi}_{st}U_{sm}U_{im}\overline{U}_{jm}\overline{U}_{rm}\ket{ik}\bra{rt},
\nonumber\\%
=\hspace{-.1cm}\tfrac{1}{\braket{\widehat{\Pi}_m^A}_\psi}\hspace{-.1cm}\left[\sum_{ir}U_{im}\overline{U}_{rm}\ket{i}\bra{r}\right]%
\hspace{-.15cm}\otimes\hspace{-.15cm}{\left[\sum_{jkst}\overline{\psi}_{st}U_{sm}\overline{U}_{jm}\psi_{jk}\ket{k}\bra{t}\right]}%
\nonumber\\%
=\tfrac{1}{\braket{\widehat{\Pi}_m^A}_\psi}\widehat{\Pi}_m%
\otimes\left[\sum_{jkst}[\widehat{W}_\psi^\dag]_{t\times s}[\widehat{\Pi}_m]_{s\times j}[\widehat{W}_\psi]_{j\times k}\ket{k}\bra{t}\right],%
\nonumber\\%
=\widehat{\Pi}_m\otimes\left[\sum_{kt}\frac{[\widehat{W}_\psi^\dag\widehat{\Pi}_m\widehat{W}_\psi]_{t\times k}}{\braket{\widehat{\Pi}_m^A}_\psi}\,\ket{k}\bra{t}\right],%
\nonumber\\%
\label{eq:apenA11}\therefore\quad%
\ket{\psi}\bra{\psi}\Big|_{\Pi_m^A}=%
\widehat{\Pi}_m\otimes\rho_{B|\Pi_m^A}.
\end{gather}
Where we have defined
\begin{equation}\label{eq:apenA12}%
\rho_{B|\Pi_m^A}=
\sum_{ij}\frac{\bra{i}\widehat{W}_\psi^\dag\widehat{\Pi}_m\widehat{W}_\psi\ket{j}}{\tr{\widehat{W}_\psi^\dag\widehat{\Pi}_m\widehat{W}_\psi}}\,\ket{j}\bra{i}%
\equiv\ket{\widehat{\psi}_B}\bra{\widehat{\psi}_B}.
\end{equation}
In the equation \eqref{eq:apenA12} it has been replace in the equation \eqref{eq:apenA9}. We can
show that \eqref{eq:apenA12} is a pure state, since it is a one-rank projector operator. A
straightforward calculation leads to
$\tr{\rho_{B|\Pi_m^A}}=1$ and $\rho_{B|\Pi_m^A}^2=\rho_{B|\Pi_m^A}$.
In the case of projective measurement in the partition $B$ the results are
similar, except for the transpose operation in the matrix
$\widehat{\mathbb{W}}_\psi$.
\\
\par\indent%
Finally, we perform a local projective measurement on the GWLs \eqref{eq:4} in
the partition $A$. According to the equation \eqref{eq:14a} we have
\begin{gather}
\rho_{\textrm{GWL}|\Pi_{m}^{A}}=\frac{(\widehat{\Pi}_{m}^{A})\rho_{\textrm{GWL}}(\widehat{\Pi}_{m}^{A})^{\dag}}{p_{m}^{A}}%
\nonumber\\
\rho_{\textrm{GWL}|\Pi_{m}^{A}}%
=\tfrac{1}{p_{m}^{A}}\widehat{\Pi}_m^A\left[\tfrac{1-p}{4}\hat{\openone}_4+p\ket{\psi}\bra{\psi}\right](\widehat{\Pi}_m^A)^\dag%
\nonumber\\
\rho_{\textrm{GWL}|\Pi_{m}^{A}}%
=\tfrac{1}{p_{m}^{A}}\left[\tfrac{1-p}{4}\widehat{\Pi}_m^A(\widehat{\Pi}_m^A)^\dag+p\widehat{\Pi}_m^A\ket{\psi}\bra{\psi}(\widehat{\Pi}_m^A)^\dag\right]%
\nonumber\\
\rho_{\textrm{GWL}|\Pi_{m}^{A}}%
=\tfrac{1}{p_{m}^{A}}\left[\tfrac{1-p}{4}\widehat{\Pi}_m^A+p\braket{\widehat{\Pi}_m^A}_\psi\ket{\psi}\bra{\psi}\Big|_{\Pi_m^A}\right]%
\nonumber\\
\label{eq:apenA14}%
\rho_{\textrm{GWL}|\Pi_{m}^{A}}%
=\widehat{\Pi}_m\otimes\left[\frac{1-p}{4p_{m}^{A}}\hat{\openone}_2+\frac{p\braket{\widehat{\Pi}_m^A}_\psi}{p_m^A}\ket{\widehat{\psi}_B}\bra{\widehat{\psi}_B}\right]%
\end{gather}
Where we have used Eqs.~\eqref{eq:apenA15}, \eqref{eq:apenA11} and \eqref{eq:apenA12}. Defining the
mixing parameter in the partition $B$ as
\begin{equation}\label{eq:apenA16}%
x_m(p)=\frac{p\braket{\widehat{\Pi}_m^A}_\psi}{p_m^A}=\frac{p\braket{\widehat{\Pi}_m^A}_\psi}{\tfrac{1-p}{2}+p\braket{\widehat{\Pi}_m^A}_\psi}.
\end{equation}
While the term that accompanies the identity matrix in \eqref{eq:14a} can be
written as
\begin{equation}\label{eq:apenA17}%
\frac{1-p}{4p_{m}^{A}}=\frac{1-p}{4p\braket{\widehat{\Pi}_m^A}_\psi/x_m(p)}=\frac{1-x_m(p)}{2}.
\end{equation}
As shown in equation \eqref{eq:14a}. This result is very important since the projective measurement
does no alter the structure of the GWLs, but modifies the mixing parameter $p$ by $x_m(p)$.

\section{Calculation of condicional entropy for Werner-like states}%
\label{sec:apenB}%
For the optimization process, it is convenient to define the equation \eqref{eq:23}, which is a
positive and monotonically increasing function of the mixing parameter $x_m(p)$, of partition $B$.
So that the conditional entropy \eqref{eq:20} is given by
\begin{equation}\label{eq:apenB1}%
S_{B|\{\Pi_m^A\}}(\psi,p)%
=\min_{\{\Pi_m^A\}}\sum_{m}F\left(x_m(p)\right).
\end{equation}
The minimum is obtained when there is a set of value for the mixing parameter $x_m(p)$ such that
the function $F$ is minimal, subject to restriction \eqref{eq:18}. For the case $n=2$, it is
sufficient to find the value $\underline{x}_0$ for which $F$ is minimal, while $\underline{x}_1$ is
obtained from \eqref{eq:18}. Deriving $F(z_m)$ with respect to $z_m$ and after a simple
calculation, we can obtain
\begin{equation}\label{eq:apenB2}%
dF(x_m)=-\frac{(1-p)\log_2\left(\tfrac{1+x_m}{2}\right)}{2(1-z_m)^2}\,dx_m.
\end{equation}
Using the values of $x_{m}$ given in \eqref{eq:22a}, it is easy to show that
\begin{equation}\label{eq:apenB3}%
dF(x_m)=p\log_2\left[\frac{\tfrac{1-p}{2}+p\braket{\widehat{\Pi}^A_m}_\psi}{\tfrac{1-p}{4}+p\braket{\widehat{\pi}^A_m}_\Psi}\right]\,d\braket{\widehat{\Pi}^A_m}_\psi.%
\end{equation}
\\*[-.5cm]
\par\noindent%
It is clear from \eqref{eq:apenB3} that the process of minimizing the
conditional entropy is relegated to finding the values of $x_{m}$ that minimize
the probability $\braket{\widehat{\Pi}^A_m}_\psi$, which in turn minimize the
function $F(x_m)$. This probability presents oscillations around the uniform
distribution, which allows us to evaluate its minimum quickly. Considering the
local projective measurement \eqref{eq:8a} and after straightforward
calculation, we obtain the simplified result
\begin{equation}\label{eq:apenB4}%
\begin{split}
\braket{\widehat{\Pi}_m^A}_\psi=\tfrac{1}{2}&\Big[1+\braket{\sigma_z}_{\rho_A(\psi)}\cos(2\theta+m\pi)%
\\%
&+\braket{\exp^{i\phi\sigma_z}\sigma_x}_{\rho_A(\psi)}\sin(2\theta+m\pi)\Big]%
\end{split}
\end{equation}
where $\rho_A(\psi)$ is given by \eqref{eq:apenA7} and the explicit expressions
for the coefficients of the trigonometric functions are
\begin{subequations}\label{eq:apenB5}%
\begin{eqnarray}
\label{eq:apenB5a}%
&&\hspace{-1cm}\braket{\sigma_z}_{\rho_A(\psi)}=\sqrt{|z_{1}|^2+|z_{2}|^2-|z_{3}|^2-|z_{4}|^2}\,,%
\\
\label{eq:apenB5b}%
&&\hspace{-1cm}\braket{\exp^{i\phi\sigma_z}\sigma_x}_{\rho_A(\psi)}=2\Re\left[(z_{1}\overline{z}_{3}-z_{2}\overline{z}_{4})\exp^{-i\phi}\right].
\end{eqnarray}
\end{subequations}
Taking into account that $\Re\left[z\exp^{i\phi}\right]\le|z|$ and
\begin{displaymath}
2|z_{1}\overline{z}_{3}-z_{2}\overline{z}_{4}|=%
\sqrt{\braket{\sigma_x}_{\rho_A(\psi)}^2+\braket{\sigma_y}_{\rho_A(\psi)}^2},
\end{displaymath}
we have that the amplitude of the oscillations presented in \eqref{eq:apenB4}
is given by
\begin{eqnarray}
A_\psi%
&=&\tfrac{1}{2}\sqrt{\braket{\sigma_z}_{\rho_A(\psi)}^2+\braket{\sigma_x}_{\rho_A(\psi)}^2+\braket{\sigma_y}_{\rho_A(\psi)}^2},
\nonumber\\
&=&\frac{1}{2}\sqrt{\sum_{i=1}^3\braket{\sigma_i}_{\rho_A(\psi)}^2}=%
\frac{1}{2}\sqrt{\sum_{i=1}^3\tr{\sigma_i\rho_A(\psi)}^2},
\nonumber
\end{eqnarray}
\begin{eqnarray}
&=&\frac{1}{2}\sqrt{\sum_{i=1}^3\tr{\sigma_i\hat{\mathbb{W}}_\psi\hat{\mathbb{W}}_\psi^\dag}^2}=
\frac{1}{2}\sqrt{\sum_{i=1}^3\tr{\hat{\mathbb{W}}_\psi^\dag\sigma_i\hat{\mathbb{W}}_\psi}^2},
\nonumber
\\%
\label{eq:apenB6}%
&=&\frac{1}{2}\sqrt{||\ket{\psi}||^4-C[\ket{\psi}]^2}=\frac{1}{2}\Delta_{\ket{\psi}}.
\end{eqnarray}
Where it has been used using the equation \eqref{eq:apenA7}.  This result coincides with
\eqref{eq:25}. So the minimum probability value is
\begin{equation}
\braket{\widehat{\Pi}_m^A}_\psi^{\min}=\tfrac{1}{2}-A_\psi=%
\tfrac{1}{2}\left(1-\Delta_{\ket{\psi}}\right).
\end{equation}
Sintetizing, the value that minimize the function $F(x_m(p))$, and therefore minimize the
conditional entropy \eqref{eq:apenB1}, is given by \eqref{eq:24a}.

\begin{small}

\end{small}

\end{document}